# Comment: Demystifying Double Robustness: A Comparison of Alternative Strategies for Estimating a Population Mean from Incomplete Data

**Greg Ridgeway and Daniel F. McCaffrey**

This article is an excellent introduction to doubly robust methods and we congratulate the authors for their thoroughness in bringing together the wide array of methods from different traditions that all share the property of being doubly robust.

Statisticians at RAND have been making extensive use of propensity score weighting in education (McCaffrey and Hamilton (2007)), policing and criminal justice (Ridgeway (2006)), drug treatment evaluation (Morral et al. (2006)), and military workforce issues (Harrell, Lim, Castaneda and Golinelli (2004)). More recently, we have been adopting doubly robust (DR) methods in these applications believing that we could achieve further bias and variance reduction. Initially, this article made us second-guess our decision. The apparently strong performance of OLS and the authors' finding that no method outperformed OLS ran counter to our intuition and experience with propensity score weighting and DR estimators. We posited two potential explanations for this. First, we suspected that the high variance reported by the authors when using propensity score weights could result from their use of standard logistic regression. Second, stronger interaction effects in the outcome regression model might favor the DR approach.

*Greg Ridgeway is Senior Statistician and Associate Director of the Safety and Justice Program at the RAND Corporation, Santa Monica, California 90407-2138, USA e-mail: gregr@rand.org. Daniel F. McCaffrey is Senior Statistician and Head of the Statistics Group at the RAND Corporation, Pittsburgh, Pennsylvania 15213, USA e-mail: danielm@rand.org.*

## 1. METHODS

We felt the authors were somewhat narrow in their discussion of weighting by focusing only on propensity scores estimated by logistic regression in their simulation. The high variability in the weights reported by the authors could result from using this method. The authors state that none of the various IPW methods could overcome the problems with estimated propensity scores near 0 and 1, yet we believed that this is indicative of a problem with the propensity score estimator rather than IPW methods. In our experience weights estimated using a generalized boosted model (GBM) following the methods of McCaffrey, Ridgeway and Morral (2004) as implemented in the Toolkit for Weighting and Analysis of Nonequivalent Groups, the twang package for R, tend not to show the extreme behavior that resulted from logistic regression (Ridgeway, McCaffrey and Morral (2006)).

GBM is a general, automated, data-adaptive algorithm that can be used with a large number of covariates to fit a nonlinear surface and estimate propensity scores. GBM uses a linear combination of a large collection of piecewise constant basis functions to construct a regression model for dichotomous outcomes. Shrunken coefficients prevent the model from overfitting. The use of piecewise constants has the effect of keeping the estimated propensity scores relatively flat at the edges of the range of the predictors, yet it still produces well-calibrated probability estimates. This reduces the risk of the spurious predicted probabilities near 0 and 1 that cause problems for propensity score weighting. Many variants of boosting have appeared in machine learning and statistics literature and Hastie, Tibshirani and Friedman (2001) provide an overview. We optimized the number of terms in the GBM model to provide the best "balance" between the weighted covariate distributions $f(\mathbf{x}|t=1)$ and $f(\mathbf{x}|t=0)$. This approach to fitting propensity scores is fully implemented in the `twang` package.

We tested our conjectures about the performance of IPW and DR estimators based on GBM and in the presence of omitted interactions terms through a simulation experiment using the same design that the authors used. Using their model from Section 1.4, we generated 1000 datasets and calculated the population and nonresponse estimator as they did for Tables 1, 3, 5 and 6. In addition, for each dataset we also estimated propensity scores using GBM (the `ps()` function optimizes the number of basis functions in the model to minimize the largest of the marginal Kolmogorov–Smirnov statistics). While their simulations do not test this, Kang and Schafer noted that choices other than logistic regression may be preferable and offered robit regression as a possibility. We included robit(1) propensity score estimates in our simulations as well.

In addition to experimenting with other propensity score estimators, we also expanded the simulation to add an interaction term equal to $20Z_1Z_2$ to the mean function for $Y$. R code for the simulation experiments is available upon request.

## 2. RESULTS

Table 1 shows the results for the IPW methods. The rows of the table correspond to the different estimators presented by Kang and Schafer. The row labeled "Model for $Y$" denotes whether the outcome OLS model used the $Z$ variables or the "mistransformed" $X$ variables in the estimation or, in the case of the interaction experiments, whether the fitted model includes a $Z_1Z_2$ term. The 12 IPW estimators in Table 1 vary by the propensity score model (logistic, GBM or robit regression), the use of $Z$ or $X$ as covariates in the propensity score model, and the use of either population weighting (IPW-POP) or the nonresponder reweighting (IPW-NR). The elements of the table contain the ratio of the RMSE of the alternative estimators to the RMSE of OLS fit with the covariates listed in each column heading.

First note that IPW estimators with logistic regression using the $X$ covariates have by far the largest RMSEs in the table. Second, while OLS seems to be preferable over IPW methods in the case where there is in truth no interaction, when the OLS models exclude an important interaction the IPW methods are preferable. When faced with the choice between OLS and IPW, the analyst must decide whether to hedge against an interaction and use IPW or choose OLS, hoping that the outcome model is specified correctly and consequently gaining a 60% improvement over GBM-based IPW or a 10% improvement over robit-based IPW.

TABLE 1
*Simulation study results for IPW methods*

| Generated data: | | | K&S model | | K&S model with interaction | | |
|---|---|---|---|---|---|---|---|
| Model for $Y$: | | | Fit with $Z$ | Fit with $X$ | Fit with $Z$ and interactions | Fit with $Z$, no interactions | Fit with $X$ |
| **OLS** | | | 1.0 (1.16) | 1.0 (1.64) | 1.0 (1.35) | 1.0 (3.58) | 1.0 (5.00) |
| Logistic | $Z$ | IPW-POP | 1.4 | 1.0 | 2.0 | 0.7 | 0.5 |
| | | IPW-NR | 1.3 | 0.9 | 1.9 | 0.7 | 0.5 |
| | $X$ | IPW-POP | 9.9 | 7.0 | 9.7 | 3.6 | 2.6 |
| | | IPW-NR | 6.0 | 4.3 | 5.9 | 2.2 | 1.6 |
| GBM | $Z$ | IPW-POP | 1.9 | 1.3 | 2.2 | 0.8 | 0.6 |
| | | IPW-NR | 1.5 | 1.0 | 2.1 | 0.8 | 0.6 |
| | $X$ | IPW-POP | 2.6 | 1.9 | 3.1 | 1.2 | 0.8 |
| | | IPW-NR | 2.2 | 1.6 | 2.7 | 1.0 | 0.7 |
| Robit | $Z$ | IPW-POP | 1.4 | 1.0 | 1.7 | 0.6 | 0.4 |
| | | IPW-NR | 1.3 | 0.9 | 2.4 | 0.9 | 0.6 |
| | $X$ | IPW-POP | 1.6 | 1.1 | 2.8 | 1.0 | 0.7 |
| | | IPW-NR | 1.6 | 1.1 | 2.9 | 1.1 | 0.8 |

The rows define the model used for the propensity score weights and the columns define the variables used in the outcome regression. The cells show the ratio of the RMSE of the estimator to the RMSE of the OLS model that used the covariates listed in the column title. The actual RMSE of the OLS model is shown in parentheses.

TABLE 2
*Simulation study results for DR methods*

| Generated data: | | | K&S model | | K&S model with interaction | | |
|---|---|---|---|---|---|---|---|
| Model for $Y$: | | | Fit with $Z$ | Fit with $X$ | Fit with $Z$ and interactions | Fit with $Z$, no interactions | Fit with $X$ |
| **OLS** | | | 1.0 (1.16) | 1.0 (1.64) | 1.0 (1.35) | 1.0 (3.58) | 1.0 (5.00) |
| Logistic | $Z$ | BC | 1.0 | 1.0 | 1.0 | 0.6 | 0.4 |
| | | WLS | 1.0 | 0.8 | 1.0 | 0.5 | 0.4 |
| | $X$ | BC | 1.0 | 51.3 | 2.6 | 85.8 | 139.2 |
| | | WLS | 1.0 | 2.0 | 1.0 | 1.1 | 1.2 |
| GBM | $X$ | BC | 1.0 | 0.9 | 1.0 | *0.6 | 0.6 |
| | | WLS | 1.0 | 0.9 | 1.0 | *0.5 | 0.6 |
| Robit | $X$ | BC | 1.0 | 1.9 | 1.0 | *0.5 | 1.2 |
| | | WLS | 1.0 | 1.5 | 1.0 | *0.5 | 1.0 |

*These estimators use $Z$ in the propensity score model.
The rows define the model used for the propensity score weights and the columns define the variables used in the outcome regression. The cells show the ratio of the RMSE of the estimator to the RMSE of the OLS model that used the covariates listed in the column title. All GBM and robit models were fit using $X$ with the exception of the "Fit with $Z$, no interactions" column for which they were fit with $Z$. The actual RMSE of the OLS model is shown in parentheses.

The aim of DR estimators is to avoid this dilemma and the associated hedging by combining the benefits of both the outcome and selection models. Kang and Schafer's results suggest that current DR estimators can disappoint us. They show DR estimators having twice the RMSE as OLS estimators when both the outcome and selection models use the $X$ covariates. We investigated this using the same propensity score models described previously and the bias corrected (BC) and weighted least squares (WLS) described by Kang and Schafer. Table 2 compares the relative efficiency of DR estimators in terms of the RMSE of the DR estimators compared to OLS. The most interesting comparisons are those for which both the propensity score model and the outcome regression model use $X$. Other combinations, such as the propensity score fit with $X$ and the outcome regression fit with $Z$, are not realistic but are included for completeness.

The results clearly show that WLS with GBM dominates OLS. When the model for $Y$ is correct, WLS is essentially as efficient as the OLS estimator. When the model for $Y$ is incorrect, WLS with GBM can be significantly more efficient than OLS. GBM also outperforms the robit regression model that the authors suggested as an option. These results suggest that DR estimators might be reliable methods of buying insurance against model misspecification without paying a high price in lost efficiency.

## 3. SUMMARY

In the simulation the doubly robust estimators are particularly useful when the model is missing an important interaction between pretreatment variables. Exploratory data analysis could be used to find such missing terms in the model and hence the advantages of WLS might appear overstated. However, such exploratory analyses require modeling the outcome and present the opportunity for the model selection to be corrupted by the impact of alternative models on the estimated treatment effect. That is, the model might be chosen because it yields significant treatment effects. This type of model fitting removes one of the benefits of the propensity score approach, which is the ability to control for pretreatment variable prior to seeing the outcome to avoid the temptation or even the appearance of data snooping.

Doubly robust estimators with GBM appear to have the desired properties in this simulation study. When the model for the mean is correct, there is no cost for using the doubly robust estimator (bias corrected or WLS). They are essentially as efficient as the correctly specified OLS model. When the OLS model is incorrect, again the doubly robust estimators are at least as efficient as OLS and substantially more efficient when the OLS model is missing important interaction terms. While it is clear that more work on these estimators is needed, our results do suggest that doubly robust estimation should not be dismissed too quickly.